# Numerical Simulations of the Random Phase Sine Gordon Model.

David J. Lancaster and Juan J. Ruiz-Lorenzo

Dipartimento di Fisica and Infn, Università di Roma *La Sapienza*

P. A. Moro 2, 00185 Roma (Italy)

`djl@liocorno.roma1.infn.it`
`ruiz@chimera.roma1.infn.it`

July, 1995

**Abstract**

We have performed comprehensive numerical simulations of the Random Phase Sine Gordon Model, studying both statics and dynamics for various values of the coupling. The glass transition can be seen both in static and dynamic signals at a temperature that depends on the coupling. Our results agree qualitatively (statics) and quantitatively (dynamics) with Renormalisation Group predictions.

# 1   Introduction

The Random Phase Sine-Gordon model has various physical interpretations of interest such as a crystalline surface with disordered substrate [1] and an array of flux lines in a superconducting film subject to random pinning and parallel magnetic field [2] . The model has recently been the scene of considerable activity, both theoretical and numerical (see for instance [3, 4, 5, 6, 7, 8, 9]), and in neither area is there agreement on the details of the low temperature phase. In this letter we present a comprehensive numerical study of both dynamic and static aspects of the model for a range of values of the interaction parameter $\lambda$. We also compare with the disordered discrete gaussian model which was the subject of earlier work [8]. This study reveals simultaneous movement of the static and dynamic critical temperatures as $\lambda$ is varied, a phenomenon not previously visible in more restricted studies.

The Hamiltonian of the Random Phase Sine-Gordon model is:

$$H = \frac{\kappa}{2} \sum_{<ij>} (\phi_i - \phi_j)^2 - \lambda \sum_i \cos(2\pi(\phi_i - \eta_i)), \qquad (1)$$

where $<ij>$ denotes nearest-neighbors, $\phi_i$ is a continuous variable and $\eta_i \in [0,1)$ is the quenched random disorder.

The discrete gaussian model with a disordered substrate is related to the $\lambda \to \infty$ limit of this model.

$$H = \frac{\kappa}{2} \sum_{<i,j>} (\phi_i - \phi_j)^2, \qquad (2)$$

where the field is now integer valued, up to shifts by the disorder: $\phi_i \equiv n_i + \eta_i$ ; $n_i \in \mathbf{Z}$.

We employ a Metropolis Monte-Carlo algorithm, and for the discrete gaussian model the proposed change in the field is naturally $\pm 1$. For the continuous model the move is chosen from a distribution consisting of three flat regions, each of width $\epsilon$ and respectively centred about 0 and $\pm 1$. This form of move is far more efficient than one taken from a single flat distribution centred at zero, and makes the connection between the two models clearer. We have fixed the relative height of each region so that about half the moves are large. The widths $\epsilon$ are adjusted at each temperature and value of $\lambda$ to give an acceptance rate of approximately 50%.

We have used the computer APE [13], simulating in parallel 256 systems composed of 128 pairs with distinct disorder (samples). Each pair of uncoupled replicas with the same disorder was subject to different thermal noise, and in the standard procedure for disordered systems the overlap between the copies was monitored. The disorder average was taken over all 256 systems.

The size of the system is always $64^2$, studies of the discrete gaussian model for larger sizes were reported in [8]. We fix $\kappa = 2$ so the critical temperature according to theory valid for small $\lambda$ is $T_C = 2/\pi$. Various values of $\lambda$ have been considered up to $\lambda = 3.0$, thermalisation becoming increasingly difficult as $\lambda$ increases. Here we present results for $\lambda = 0.5, 2.0$.

In all cases the energy relaxes almost immediately to its final value and we therefore conclude that local equilibrium is achieved quickly, leaving the long time dynamics to be that of domain readjustment. A direct visual inspection of the configurations confirms this but a detailed investigation of the process requires the correlation function defined by:

$$C(r) = \overline{< (\phi_r - \phi_0)^2 >}, \qquad (3)$$

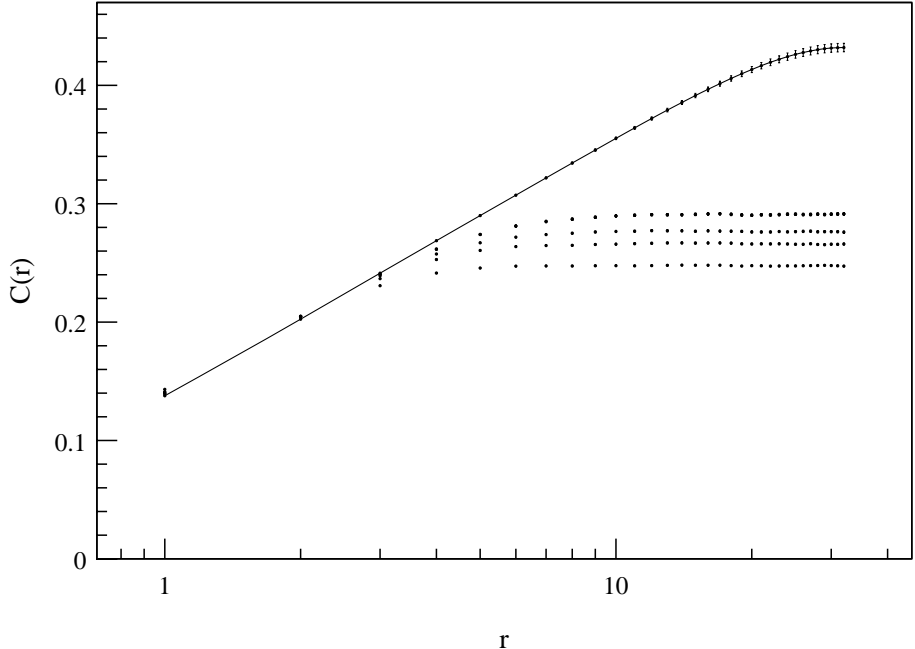

Figure 1: Correlation functions, $C_t(r)$, shown at equal time intervals (bottom to top: $t$ =2000, 4000, 6000 and 8000 sweeps), and the equilibrium correlator, $C_{\text{asy}}(r)$, as the continuous line ($\lambda = 2.0$, $T = 0.5$).

where $\langle \, \cdot \, \rangle$ and $\overline{\phantom{..}}$ denote thermal and disorder averages respectively.

## 2  Dynamics

It is convenient to plot the correlator against $\log r$ as in figure 1, the top curve represents the equilibrium form. This equilibrium curve is fairly linear except for the finite size effects and a small component of a different large distance behavior which will be the subject of the next section. The other curves in figure 1 illustrate the approach to this final equilibrium form, the short distance part of the correlators follows the equilibrium curve but a plateau is reached beyond a distance that can intuitively be identified as the correlation length $\xi(t)$ at that time.

This intuitive definition of the time dependent correlation length is numerically robust since the plateaus are very well defined. Knowing the asymptotic form of the correlator ($C_{\text{asy}}(r)$), it is simple to obtain the correlation length $\xi(t)$ from $C_{\text{asy}}(\xi(t)) = \text{Plateau}[C_t(r)]$, where $C_t(r)$ is the correlation function at the time $t$. Starting from a constant configuration (i.e. $\phi_i = \text{Const.}$), we measure how the correlation length grows with time. Provided $\xi(t)$ does not approach the system size, we find that dynamical scaling is well obeyed:

$$\xi(t) \simeq t^{1/z}. \tag{4}$$

We show this growth in figure 2 on a log-log scale. The errors on each point arise from variations between different samples. We emphasise that this scaling behavior is independent of the detailed choice of local Monte-Carlo dynamics. We have checked this explicitly by comparing with the slower dynamics based on a simple move taken from a single flat distribution centred around zero, and in fact have used this alternative move for our determinations of $z$. The dependence of $z$ on temperature is shown in figure 3 for $\lambda = 0.5, 2.0$ and $\infty$. At high temperatures the dynamic exponent takes its gaussian value of 2, marked by a line in the figure. Below, but in the vicinity of the transition the deviation from the gaussian value appears to be linear in the reduced temperature. This observation would be in accord with the RG prediction for $T < T_C$ [10]

$$z = 2 + 2e^{\gamma}(1 - T/T_C), \qquad (5)$$

where the formula is valid for small $\lambda$ with $T_C = 2/\pi$ and $\gamma$ is Euler's constant. If the linearity continues to hold for large values of $\lambda$ we can determine transition temperatures reasonably accurately and it is clear that these depend on $\lambda$. In any event, the large $\lambda$ data is certainly not compatible with $T_C = 2/\pi$. The formula (5) yields a $\lambda$ independent slope with numerical value 3.56. We can try to extract the slope and critical temperature assuming a fit of the form

$$z = 2 + a_D(1 - T/T_D). \qquad (6)$$

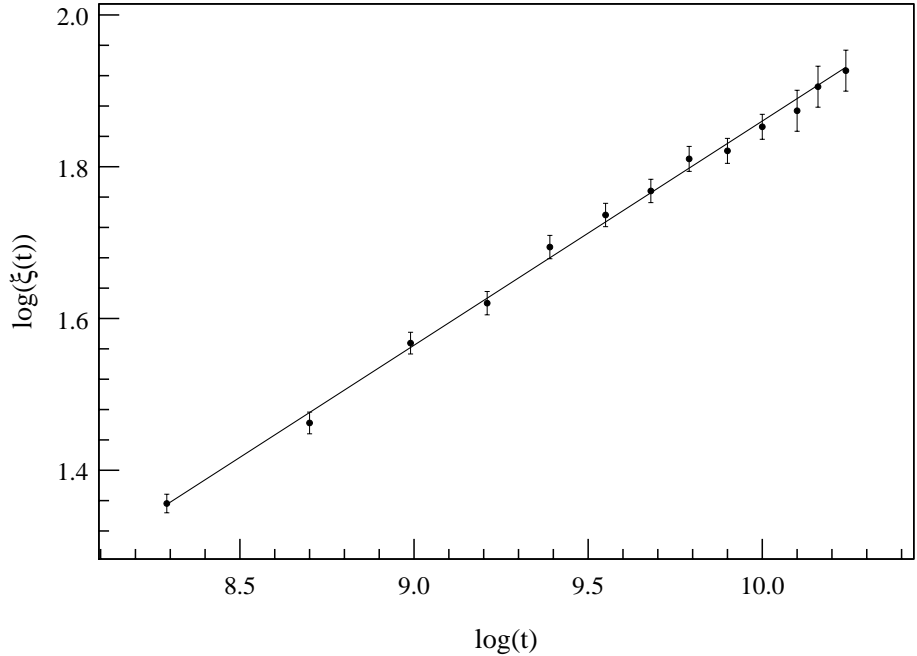

Figure 2: Growth of correlation length with time ($\lambda = 2.0, T = 0.5$).

The results for $a_D$ and $T_D$ are reported in Table 1 (left). Different sets of points have been used to demonstrate that the fit is stable, but we always ignore points within twice their error of the gaussian value ($z = 2$).

An alternative type of comparison is available with results for the dynamical exponent based on the response to a small driving force $F$. In the low temperature phase RG arguments give the relation between applied force and resulting velocity as [3]: $\overline{\langle \partial_t \phi \rangle} \sim \tau^{2z} F^{1+2z\tau}$, where $\tau$ is the reduced temperature. It is reassuring that our data are in good numerical agreement, both at small [5] and large [9] $\lambda$, with such a different approach based on Langevin dynamics.

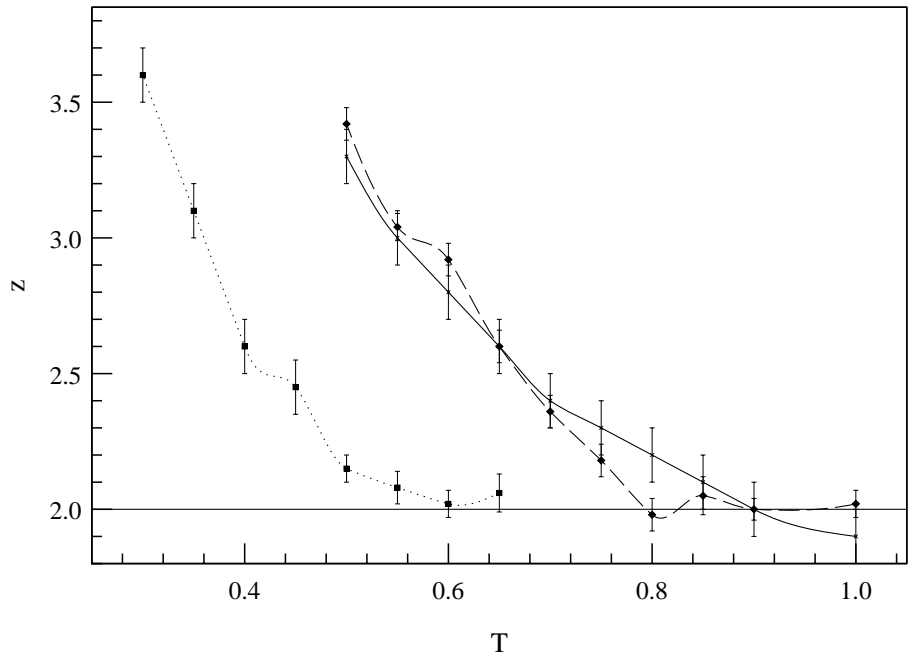

Figure 3: The dynamical critical exponent, $z$, against temperature for $\lambda = 0.5, 2.0$ and $\infty$. The gaussian value is marked with a horizontal line. We use squares and a dotted line for $\lambda = 0.5$, diamonds and a dashed line for $\lambda = 2.0$ and crosses and a continuous line for $\lambda = \infty$. The lines are only to guide the eye.

|   | Dynamics | | | Statics | | |
|---|---|---|---|---|---|---|
| $\lambda$ | Interval | $a_D$ | $T_D$ | Interval | $a_S$ | $T_S$ |
| 0.5 | [0.4, 0.5] | 2.5(5) | 0.53(2) | [0.4, 0.5] | 0.69(12) | 0.58(2) |
|  | [0.35, 0.5] | 3.1(3) | 0.52(1) | [0.35, 0.5] | 0.67(7) | 0.58(1) |
| 2.0 | [0.5, 0.75] | 3.8(2) | 0.78(1) | [0.5, 0.75] | 0.54(6) | 0.77(2) |
|  | [0.6, 0.7] | 4.3(6) | 0.76(2) | [0.55, 0.75] | 0.44(12) | 0.78(4) |
| $\infty$ | [0.55, 0.8] | 2.7(3) | 0.85(3) | [0.5, 0.80] | 0.44(7) | 0.87(2) |
|  | [0.6, 0.75] | 2.8(6) | 0.83(4) | [0.6, 0.80] | 0.35(5) | 0.90(2) |

Table 1: Results of linear fit to $z(T)$ (dynamics) and quadratic fit to $b_2(T)$ (statics) for three values of $\lambda$ and for different sets of temperatures (intervals). See text for notation. All the fits in this table have $\chi^2/\text{d.o.f} \approx 1$.

We have shown in this section that in the vicinity of the critical temperature the $z$-exponent has a linear dependence on reduced temperature but that the critical temperature depends on $\lambda$. The slope is in tolerable agreement with the prediction from RG.

## 3 Statics

In this section we discuss the equilibrium form of the correlation function along the lines of previous work on the discrete gaussian model [8] and shall repeat some of the arguments used there. The main priority to to ensure that thermal equilibrium is actually obtained. An annealing scheme is used and it is important to include the large Monte-Carlo moves discussed above, even so, very long times are needed. Typically for each temperature, we thermalise for $10^5$ sweeps, measure over a period of $2.10^4$ sweeps, and repeat this cycle several times. The system is taken to be thermalised when we reach a situation in which subsequent measuring cycles show no systematic drift and fluctuations within each measurement cycle are similar to those between cycles.

At high temperatures the correlator is accurately fitted by a single log, or more precisely by the lattice version:

$$P_L(r) = \frac{1}{2L^2} \sum_{n_1=1}^{L-1} \sum_{n_2=0}^{L-1} \frac{1 - \cos(\frac{2\pi r n_1}{L})}{2 - \cos(\frac{2\pi n_1}{L}) - \cos(\frac{2\pi n_2}{L})} \simeq \frac{1}{2\pi} \log(\frac{r}{2\sqrt{2}e^\gamma}) , \qquad (7)$$

where the symbol $\simeq$ holds for $L \gg 1$. The coefficient follows the temperature to within 1%, in agreement with all theoretical predictions [4, 11].

As the temperature is reduced below some value this fit becomes substantially worse. The degradation cannot be ascribed to short distance lattice effects since a similar worsening occurs for fits in which short distance points are omitted. We emphasise that the long distance finite size effects are well under control in equation (7). Our results are not sensitive enough to determine the functional form of the correction necessary but theoretical prejudice suggests a fit of the form [1]:

$$C(r) = b_1 P_L(r) + b_2 P_L(r)^2. \qquad (8)$$

This fit works very well for all lambda, improving as equilibrium is approached, and with a value of $\chi^2$ that is small and does not vary significantly throughout the temperature range. The fit is made by an exact minimisation procedure and errors are determined by jack-knife. Figures 4 and 5 show the resulting values of the coefficients for $\lambda = 0.5$ and $2.0$, and also the data for $\lambda = \infty$ from reference [8].

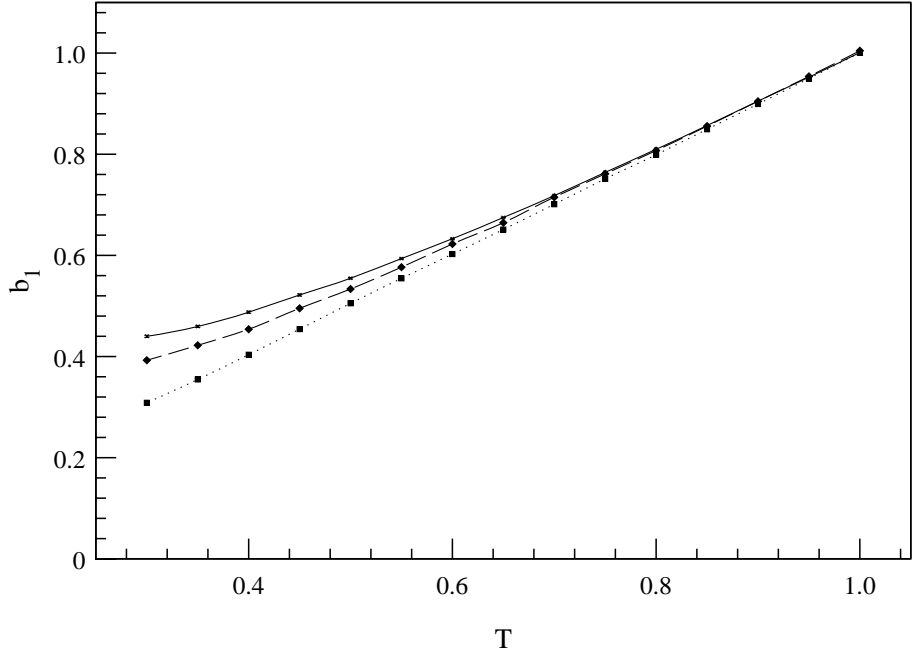

Figure 4: Coefficient of the log ($b_1$) versus the temperature for different values of the $\lambda$ parameter from a two parameter fit. The same symbols as in figure 3. The error bars are smaller that the size of the points.

The first coefficient $b_1$ continues to follow its high temperature behavior down to low temperatures showing only small deviations from linearity. The second coefficient $b_2$ is sensitive to the transition and grows at low temperature, it does not however display the same kind of linearity evident in the dynamic case. It is therefore slightly more difficult to determine the critical temperature though it is clear that this is $\lambda$ dependent.

The form (8) is the result of RG calculations [1] which predict a universal behavior for the second coefficient $b_2 = 8\tau^2 + O(\tau^3)$ [2]. This quadratic dependence on reduced temperature can be tested, and we show in Table 1 (right) the results of a fit

$$b_2 = a_S(1 - T/T_S)^2, \qquad (9)$$

where we have only used points for which $b_2$ is further than twice the error away from zero. The fit is good, but in this case the value of the coefficient is not even close to the RG prediction.

The transition identified in the dynamic and static measurements should occur at the same temperature. This follows from the RG analysis and also from a variational calculation of the entropy of metastable states [14, 15]. We see from Table 1 that both static and dynamic measurements give similar critical temperatures, and most importantly, that these shift simultaneusly with different values of $\lambda$.

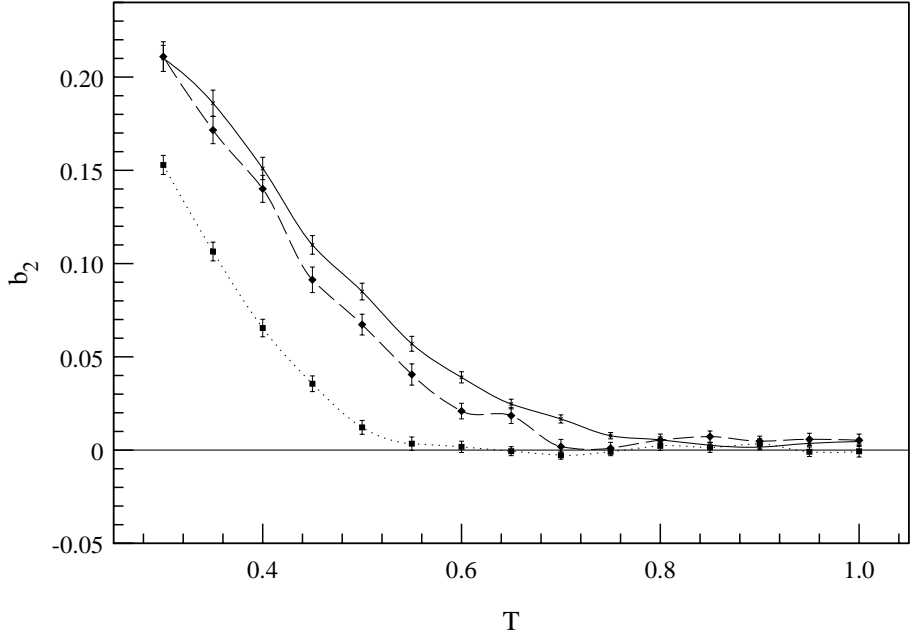

Figure 5: Coefficient of the $\log^2$ ($b_2$) versus the temperature for different values of the $\lambda$ parameter from a two parameter fit. The same symbols as in figure 3.

## 4 Conclusions

This comprehensive study reveals clear signals of a transition both in the static and dynamic properties of the system. The transition seems to occur at the same temperature whether determined by static or dynamic methods. The transition temperature does however depend on $\lambda$, being close to $\kappa/\pi$ only at small $\lambda$. We find the general behavior predicted by RG arguments, that is: linear behavior of the dynamic exponent, $z \propto \tau$, and quadratic behavior of the coefficient of $\log^2$, $b_2 \propto \tau^2$. Detailed numerical agreement of the coefficients, especially in the case of the quadratic behavior, is lacking (by a factor of $\sim 14$) while in the dynamic case the agreement is reasonable. Finally, an explanation of the $\lambda$ dependence of all the quantities measured is needed.

The essential difference of opinion on the low temperature phase is between analyses based on the RG that predict a $\log^2$ form for the correlator, and analysis based on a variational approach that yield a log. The RG approach has been criticised because the solutions are unstable with respect to breaking of replica symmetry [12], however the significance of this is not clear since the techniques for dealing with replicas are not sufficiently developed to deal effectively with non mean field situations. On the other hand, the variational approach has only been calculated for a gaussian ansatz which corresponds to leading order in some $1/N$ expansion. Recent results [16] using RG arguments for an N-component version of the Random Phase Sine Gordon model are interesting in that they calculate the coefficient of the $\log^2$ term to be order $1/N^3$. We hope that further analytic work in both approaches to understand the smallness of the $\log^2$ coefficient seen in this and all other simulations, will lead to a resolution of the puzzle.

Finally we believe that we have shown that it is possible obtain reliable and accurate numerical data for this rather contentious subject that can be extended and used to further test analytical predictions [17].

# Acknowledgments

We acknowledge useful discussions with B. Coluzzi, E. Marinari, G. Parisi, M. Potters and R. Monasson. J. J. R.-L. is supported by a MEC grant (Spain), D. L. by an EC HCM grant.